\documentclass{elsart}
\usepackage{yjsco}
\usepackage{amsmath}
\usepackage{fancyvrb}
\usepackage[numbers,sort&compress]{natbib}
\usepackage{hypernat}
\usepackage{hyperref}
\usepackage{graphicx}
\usepackage{xspace}
\newcommand{\bs}{$\backslash$}
\newcommand{\Cpp}{\leavevmode\rm{\hbox{C\hskip -0.1ex\raise 0.5ex\hbox{\tiny ++}}}\xspace}

\makeatletter
\def\ps@copyright{}
\makeatother

\fvset{frame=lines,framerule=0.1pt,framesep=8pt,numbers=none,xleftmargin=5ex,fontfamily=tt,fontsize=\small}
\newenvironment{screen}{\vspace{1ex}\Verbatim}{\endVerbatim\vspace{1ex}}

\begin{document}
\begin{flushright}
AEI-2006-037\\cs.SC/0608005
\end{flushright}
\begin{frontmatter}
\title{A field-theory motivated approach to symbolic computer algebra}
\author{Kasper Peeters}
\address{Max-Planck-Institut f\"ur Gravitationsphysik, Albert-Einstein-Institut\\Am M\"uhlenberg
  1, 14476 Golm, GERMANY}
\ead{kasper.peeters@aei.mpg.de}
\begin{abstract}
  Field theory is an area in physics with a deceptively compact
  notation. Although general purpose computer algebra systems, built
  around generic list-based data structures, can be used to represent
  and manipulate field-theory expressions, this often leads to
  cumbersome input formats, unexpected side-effects, or the need for a
  lot of special-purpose code. This makes a direct translation of
  problems from paper to computer and back needlessly time-consuming
  and error-prone.  A prototype computer algebra system is presented
  which features \TeX-like input, graph data structures, lists with
  Young-tableaux symmetries and a multiple-inheritance property
  system. The usefulness of this approach is illustrated with a number
  of explicit field-theory problems.
\end{abstract}
\end{frontmatter}


\section{Field theory versus general-purpose computer algebra}

For good reasons, the area of general-purpose computer algebra programs
has historically been dominated by what one could call ``list-based''
systems. These are systems which are centred on the idea that, at the
lowest level, mathematical expressions are nothing else but nested
lists (or equivalently: nested functions, trees, directed acyclic
graphs,~\ldots). There is no doubt that a lot of mathematics indeed
maps elegantly to problems concerning the manipulation of nested
lists, as the success of a large class of LISP-based computer algebra
systems illustrates (either implemented in LISP itself or in another
language with appropriate list data structures).  However, there are
certain problems for which a pure list-based approach may not be the
most elegant, efficient or robust one.

That a pure list-based approach does not necessarily lead to the
fastest algorithms is of course well-known. For e.g.~polynomial
manipulation, there exists a multitude of other representations which
are often more appropriate for the problem at hand. An area for which
the limits of a pure list-based approach have received less attention
consists of what one might call ``field theory'' problems.  Without
attempting to define this term rigorously, one can have in mind
problems such as the manipulation of Lagrangians, field equations or
symmetry algebras; the examples discussed later will define the class
of problems more explicitly. The standard physics notation in this
field is deceptively compact, and as a result it is easy to overlook
the amount of information that is being manipulated when one handles
these problems with pencil and paper. As a consequence, problems such
as deriving the equations of motion from an action, or verifying
supersymmetry or BRST invariance, often become a tedious transcription
exercise when one attempts to do them with existing general-purpose
computer algebra systems. Here, the inadequateness of simple lists is
not so much that it leads to sub-optimal, slow solutions (although
this certainly also plays a role at some stage), but rather that it
prevents solutions from being developed at~all.

To make the problems more concrete, let us consider a totally
arbitrary example of the type of expressions which appear in field
theory.  A typical Lagrangian or Noether charge might contain terms of
the type
\begin{equation}
\label{e:ex1}
\int\!{\rm d}^nx\, \frac{\partial}{\partial x^\lambda} \Big(F_{\mu\nu\rho}\, \psi^{\mu} \Big)\, \psi^{\nu}\,.
\end{equation}
Let us take, purely as an example, ~$F_{\mu\nu\rho}$ to be a
commuting field strength of some two-form field, $\psi^\mu$ an
anti-commuting vector, and $x^\mu$ to label an~$n$-dimensional space.
Traditionally, one would represent~\eqref{e:ex1} in the computer as a
nested list, which in tree-form would take the form\vspace{2ex}
\begin{center}
\includegraphics[height=5cm]{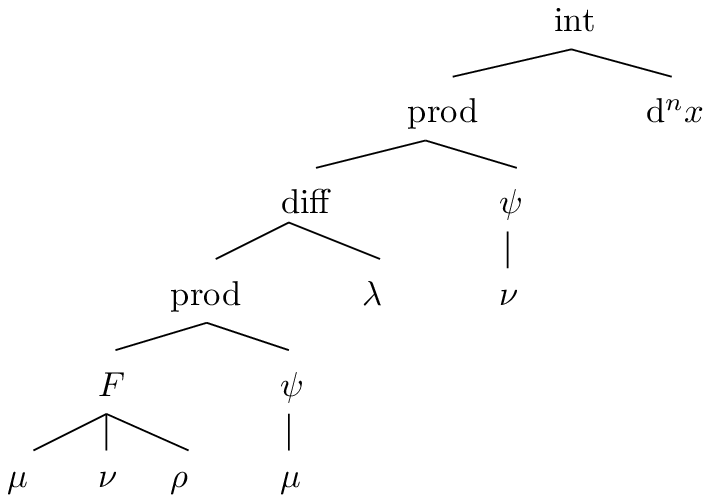}\\[2ex]
\end{center}
The precise details do not matter here; the important point is that
the expression takes the form of a multiply nested list. However, this
list can only be the starting point, since expression~\eqref{e:ex1} clearly
contains much more information than just the tree structure. This
leads to a number of ``standard'' problems which field-theory computer
algebra systems have to face: \medskip

\begin{itemize}
\item The names of contracted indices do not matter, in fact, it is
  only the contraction which is relevant. The nested list structure
  does not capture the cyclic graph-structure inherent in the
  expression. How do we make a list-based program aware of this fact,
  especially when doing substitutions? (this problem is more
  colloquially known as the ``dummy index'' problem).
\item The expression is, for the ``outside world'', an object with two
  free indices, $\lambda$ and $\rho$. However, these indices, or graph
  edges, occur at different depths in the tree. How can we access the
  nested list by the free edges of the tree?
\item The reason why $F$ and $\psi$ should stay as children of the
  diff node is that they depend on~$x$. Where do we store this
  information? And how do we make algorithms aware of~it?
\item The diff and $\psi$ child nodes of the prod node cannot be moved
  through each other, because the diff node contains a $\psi$. In
  other words, the diff node ``inherits'' the anti-commutativity from
  one of its child nodes. How do we handle this?
\item The anti-symmetry of~$F$ relates several contraction patterns.
  However, there are more complicated symmetries present in the
  expression. The Bianchi identity on the field strength, for
  instance, is a multi-term relation involving the indices
  on~$F$ and the index on diff. How do we make the program aware of
  this identity, and how do we take it into account when reducing
  expressions to a canonical form?
\item For physicists, the simplest way to write expressions such
  as~\eqref{e:ex1} is to use \TeX{} notation, for example
\begin{equation*}
\verb|\int d^nx \partial_{\lambda} ( F_{\mu\nu\rho} \psi^{\mu} ) \psi^{\nu}|
\end{equation*}
Being able to input expressions like this would eliminate a large
number of errors in transcribing physics problems to computer algebra
systems, and make it much easier to use the computer as a scratch pad
for tedious calculations (in particular, it would eliminate a good
part of the learning curve of the program). Although \TeX{} notation
certainly lacks the semantics to be used as input language for a
generic computer algebra system (see e.g.~\cite{fateman99parsing} for
a discussion of this problem), it is not hard to come up with a subset
of \TeX{} notation which is both easily understandable and
mathematically unambiguous. But how do we teach an existing general
purpose system to deal with input of this type?
\end{itemize}
\medskip

This collection of problems suggest that a general-purpose system
based on ``nested lists'' is a rather bare-bones tool to describe
field-theory problems. The nested list is just one of the many
possible views or representations of the (rather heavily labelled)
graph structure representing the expression.  While it is perfectly
possible to tackle many of the problems mentioned above in a
list-based system (as several tensor algebra packages for general
purpose computer algebra systems
illustrate~\cite{e_xact,e_balf1,Klioner:1997sv,parker1}), this may not
be the most elegant, efficient or robust approach (the lack of a
system which is able to solve all of the sample problems in
section~\ref{s:examples} in an elegant way exemplifies this point). By
endowing the core of the computer algebra system with data structures
which are more appropriate for the storage of field-theory
expressions, it becomes much simpler to write a computer algebra
system which can resolve the problems listed above.\footnote{There are
of course \emph{other subclasses} of field-theory problems for which
some of the points raised here are irrelevant and efficient computer
algebra systems have been developed; see e.g.~\cite{Vermaseren:2000nd}
for an example.}

The remainder of this paper describes the key ingredients in an
approach taken in the prototype computer algebra system ``cadabra''.
Full details of this program, including source code, binaries and a
reference manual, can be found at the web site~\cite{kas_cdb}.

\section{Design goals and implementation}
\label{s:structure}

This section describes in more detail the main features of the cadabra
program: the internal graph views of the tree structure, its handling
of node symmetries and the use of a multiple-inheritance property
system.  In addition to these features, cadabra also has a number of
other characteristics which make it especially tuned to the
manipulation of ``field theory'' problems.  An important
characteristic which should not remain unmentioned is the fact that
cadabra accepts \TeX{}-like notation for tensorial expressions,
making it much easier to transcribe problems from and to paper. The
program can be used both from a text-based command-line interface as
well as from a graphical front-end or from within
\TeX{}macs~\cite{vdH:Gut}. I will not discuss the \TeX{} input/output
in detail, but examples can be found in section~\ref{s:examples}.

\subsection{Graph structure}
\label{s:graph}

Cadabra is a standalone program written in~\Cpp.  As in most other
computer algebra systems, the internal data storage used in cadabra is
that of a tree. Concretely, this is implemented using a custom tree
container class~\cite{kas_tree} based on STL ideas~\cite{b_muss1}.
However, what sets the program apart from other systems is that a)~the
tree structure contains more data than usual, to make it easier to
represent field-theory problems in a compact way, and b)~the tree
manipulation algorithms offer several ways of viewing and
modifying this tree. In more detail: \medskip

\begin{itemize}
\item The nodes of the tree carry so-called ``parent-relation''
  information, which determine how child nodes are related to parent
  nodes. As an example of such a relation, consider the
  expression~$T^{\mu}{}_{\nu}(x)$. This is stored as a node~$T$, with
  three children~$\mu$, $\nu$ and $x$, which have parent relations
  ``superscript'', ``subscript'' and ``argument'' respectively (more
  relations can easily be added in the future if necessary). A common
  way of storing this information is e.g.~\mbox{\tt T[mu, -nu, x]} or
  \mbox{\tt T[up[mu], dn[nu], x]}, but both have their disadvantages:
  the first form does not allow us to store~$T^{-\mu}{}_{\nu}(x)$,
  while the second form introduces an additional layer of overhead. A
  format similar to the second case is also used in
  Stensor~\cite{maccallum1}, albeit using a different syntax; it has
  the disadvantage that it is neither a very convenient representation
  for the computer (especially not when the implementation is in \Cpp),
  nor a representation which is convenient for the user, as it is quite
  distinct from \TeX{} notation.
\item The tree class not only provides a way to access the nodes of
  the graph by pre- or post-order traversal, but also
  provides iterators which access e.g.~only all indices of a node. In
  the example~\eqref{e:ex1}, an index iterator acting on the {\tt
  diff} node would return, in turn, the~$\mu$, $\nu$, $\rho$, $\mu$
  and $\lambda$ indices. This makes it easy to write fast
  low-level routines which deal directly with the tensor structure of
  an expression.
\item The tree manipulation algorithms are aware of the meaning of
  ``contracted nodes'' (contracted indices). Whenever one expression
  graph is inserted into another one, the algorithms automatically ensure
  that labels which are used to denote contracted indices (i.e.~edges
  which connect two nodes) are relabelled appropriately. Names are
  chosen by using property lists (see section~\ref{s:properties}).
\item The contraction detection mechanism can deal with sub- or
  superscripts which do not denote indices, as in
  e.g.~$A^\dagger$. This is achieved by attaching a special property
  to the symbol (see section~\ref{s:properties} for more details).
\end{itemize}
\medskip

The enhanced tree structure can be modified at the level of the user
interface through standard list manipulation commands, or at the level
of custom modules written in~\Cpp.

\subsection{Symmetries}
\label{s:symmetries}

A second issue which cadabra addresses differently from other computer
algebra systems is that of node symmetries. It is common in computer
algebra systems that there is a generic way to specify so-called
\emph{mono-term} symmetries. These are symmetries which relate one
particular ordering of arguments to another one, up to a possible
overall sign. Cadabra can of course find canonical representations for
tensors with mono-term symmetries (using an external
implementation~\cite{e_xact} of the double-coset
algorithm~\cite{port2}; an alternative backtracking algorithm for
mono-term symmetries is described in~\cite{dres1}).

However, mono-term symmetries do not exhaust the symmetries of tensors
transforming in generic representations of the Lorentz group. They do
not include Garnir symmetries of Young tableaux. Examples of such
symmetries are the Ricci identity for Riemann tensors,
\begin{equation}
R_{m n p q} + R_{m p q n} + R_{m q n p} = 0\,,
\end{equation}
or the Bianchi identity for field strengths. These identities relate
more than two terms, and are therefore also called \emph{multi-term}
symmetries. Despite the clear importance of being able to take such
identities into account, there are very few computer algebra systems
which have implemented a solution. The implementation
in~\cite{parker1} simply uses a large set of transformation rules for
Riemann tensor monomials. These rules were constructed by hand, and
are only available for Riemann tensor monomials up to third order
(i.e.~it would require tedious work to construct such rules for more
general expressions, involving more than just the Riemann or Ricci
tensors). An alternative approach is taken
in~\cite{maccallum1,ilyi1,e_balf1}, in which the set of all identities
for a particular tensor is used to rewrite a given expression in
canonical form. This idea of handling multi-term symmetries using a
sum-substitution algorithm goes back to at least~\cite{hornf1}.

Cadabra, instead, uses Young projector methods internally for all
index symmetry handling. The underlying idea is that by applying a
Young projector to a tensor, its multi-term symmetries become
manifest. This allows one to construct a basis of tensor monomials
constructed from arbitrary tensors, and to decompose a given monomial
on any preferred basis.\footnote{In order to determine the number of
terms in a basis of monomials of tensors, cadabra relies on an
external program for the computation of tensor product representations
(using the LiE program~\cite{e_cohe1}).} This method was first
described in~\cite{Green:2005qr}. For e.g.~Riemann tensors, the idea
is to replace all tensors by their equivalent form
\begin{equation}
\label{e:Rproj}
R_{a b c d} \rightarrow 
 \tfrac{1}{3}\big( 2\, R_{a b c d} - R_{a d b c} + R_{a c b d} \big)\,.
\end{equation}
The expression on the right-hand side manifestly satisfies the cyclic
Ricci identity, even if one only knows about the mono-term symmetries
of the Riemann tensor. Using the projector~\eqref{e:Rproj} it is easy
to show e.g.~that~$2\,R_{a b c d} R_{a c b d} = R_{a b c d} R_{a b c
d}$. The monomial on the left-hand side maps to
\begin{equation}
\begin{aligned}
R_{a b c d} R_{a c b d} \rightarrow \tfrac{1}{3} \big( R_{a b c d} R_{a c b d}
 + R_{a b c d} R_{a b c d} \big)\,,
\end{aligned}
\end{equation}
while $R_{a b c d} R_{a b c d}$ maps to twice this expression, thereby
proving the identity.

Writing each term in a sum in a canonical form by
using~\eqref{e:Rproj} would typically lead to extremely large
expressions, and not be very convenient for subsequent
calculations. However, the same algorithm can also be used to write a
sum in a ``minimal'' form.\footnote{``Minimal'' here does not
necessarily mean that the expression has been reduced to the shortest
possible form, which is a problem which to the best of my knowledge
remains unresolved. That is, while the algorithm removes dependent
terms, as in $2\,R_{a b c d} + 2\,R_{b c a d} + R_{c a b d} \rightarrow
R_{a b c d} + R_{b c a d}$ (because the third term is found to be
expressible as a linear combination of the first two), it does not
reduce this further to $- R_{c a b d}$ (typical cases are of course
more complicated than this example).} That is, by projecting each
term using~\eqref{e:Rproj} the program can perform the simplification
\begin{equation}
R_{a b c d} R_{a c b d} + R_{a b c d} R_{a b c d}
 \rightarrow 3\,R_{a b c d} R_{a c b d}\,,
\end{equation}
i.e.~express the second term in terms of the first one.  This does not
define a canonical form (the expression could equally well have been
written using~$R_{a b c d} R_{a b c d}$), but it does systematically
eliminate terms which can be written as linear combinations of other
terms.

\subsection{Properties}
\label{s:properties}

A third problem for which cadabra takes a different approach from
other systems is that of ``typing'' of symbols and expressions. In
cadabra, the meaning of symbols or expressions is determined by
associating \emph{properties} to them. Properties can be simple, such
as ``being an integer'', or ``being anti-symmetric in all indices'',
or ``being an index which is not to be summed over'' (cf.~the
discussion in section~\ref{s:graph}). They can also be more complicated
and composite, such as ``being an anti-commuting spinor in the
left-handed Weyl representation of the eight-dimensional Lorentz
group''.

The general problem of deducing properties of composite objects from
the properties of their constituents is a hard (see
e.g~\cite{weib1}). Cadabra takes a pragmatic approach, trying to
provide a useful property system for concrete problems rather than
trying to be complete or mathematically rigorous. Properties are
implemented as a standard multiple-inheritance tree of \Cpp
objects. The association to symbols is stored in a map, which relates
patterns to pointers to property objects.\footnote{It is important
that such properties are implemented at a low level. Most computer
algebra systems would allow one to implement e.g.~handling of sets of
non-commuting objects using user-defined property testing functions
and appropriate transformation rules. It is a much harder problem to
make sure that all routines of the underlying system use these
properties efficiently and correctly.}  This makes it relatively easy
to make properties inherit from each other.  An example of an
inherited property is the property {\tt PartialDerivative}, which
inherits from {\tt TableauBase}, so that the symmetry information of
objects on which a partial derivative acts are automatically
propagated.

Nodes can inherit properties from child nodes. A simple situation in
which this is useful is for instance when one uses accents to mark
symbols, as in e.g.~$\bar{\psi} \psi$. If {\tt \bs{}psi} is declared
to be self-anticommuting, we obviously want the~{\tt
\bs{}bar\{\bs{}psi\}} tree to have this property as well. When
scanning for properties of nodes, the internal algorithms take into
account such inheritance of properties. Inheritance of a property is,
itself, again implemented as a property (in the example above, the
{\tt \bs{}bar} node is declared to have the property {\tt
PropertyInherit}, while more fine-tuned inheritance is implemented by
deriving from a templated {\tt Inherit} class, as in e.g.~{\tt
Inherit<Spinor>}).\footnote{This is similar to Macsyma's types
and features: the property which is attached to a symbol is like a
`type', while all properties which the symbol inherits from child
nodes are like `features'. Property inheritance can also be found
other systems, e.g.~Axiom~\cite{daly1}.}

Not all property inheritance is, however, as simple as propagating the
information up from a deeper lying node. More complicated property
inheritance occurs when nodes have to ``compute'' their properties
from properties of the child nodes. This occurs for instance when we
want to know how products of symbols commute among each other. For
such cases, there are more complicated property classes, for instance
{\tt CommutingAsProduct} or {\tt CommutingAsSum}. Similarly, there is
a property {\tt IndexInherit} which indicate that nodes should make
the indices of their child nodes visible to the outside world. Other
composite property objects can easily be added to the system.

\section{Typical examples}
\label{s:examples}

In this section, the three main points discussed in the previous
section main text (enhanced tree data structures \& algorithms, the
use of representation theory to classify object symmetries, and the
use of properties) will be illustrated with a number of explicit
examples. These examples are meant to be readable without further
information about the program language. As such, they also illustrate
the ease with which tensorial expressions can be fed into the program.
Full details of the input language and transformation algorithms can
be found in the manual~\cite{kas_cdb}.

\subsection{Index handling and substitution}

When doing computations by hand, we do index relabelling almost
automatically when a clash occurs. However, unless the computer
program is aware of this problem at a low level, clashes are bound to
occur frequently. Consider first the standard type of relabelling,
illustrated by the expressions
\begin{equation}
C = A^2\,,\quad \text{with}\quad A = B_{m n} B_{m n}\quad\text{and}\quad
B_{n p} = T_{m n} T_{m p}\,.
\end{equation}
In cadabra one can e.g.~do\footnote{As alluded to in the first
  section, the notation used here is not generic~\TeX{} but rather a
  well-defined subset, with some additional conventions required to
  make the input unambiguous. An example of such a convention is the
  use of spaces to separate indices; further details about the input
  format conventions can be found in the reference
  manual~\cite{kas_cdb}.}
\begin{screen}
{m,n,p,q#}::Indices(vector).
C:= A A;
@substitute!(
@substitute!(
\end{screen}
where the meaning of the hash symbol on the declaration of the~$q$
index (in the first line) will become clear soon. The result is
\begin{screen}
C:= T_{q2 m} T_{q2 n} T_{q3 m} T_{q3 n} T_{q4 p} T_{q4 q1} T_{q5 p} T_{q5 q1};
\end{screen}
This type of relabelling and automatic index generation is not an
entirely uncommon feature to find in tensor algebra systems, although
it is often implemented in an add-on package. The situation becomes
more complicated when we have indices which do not occur at the same
level, for instance
\begin{equation}
  C = A^2\,,\quad \text{with} \quad
  A = \partial_m (B_n B_p + C_{n p} ) B_{m n p}\quad\text{and}\quad
  B_n = T_{n m} S_{m}\,.
\end{equation}
Few systems know how to deal with these types of expressions
elegantly (i.e.~without requiring a cumbersome input format).  The
reason is that the derivative carries an index, but the objects in the
product on which it acts carry indices too, and these indices do not
all occur at the same depth of the expression tree. The cadabra
instructions, however, remain equally simple as in the previous example,
\begin{screen}
{m,n,p,q#}::Indices(vector).
\partial{#}::Derivative.
C:= A A;
@substitute!(
@substitute!(
\end{screen}
The result comes out as the expected
\begin{screen}
C:= \partial_{m}(T_{n q4} S_{q4} T_{p q5} S_{q5} + C_{n p}) B_{m n p} 
    \partial_{q1}(T_{q2 q6} S_{q6} T_{q3 q7} S_{q7} + C_{q2 q3}) B_{q1 q2 q3};
\end{screen}
Finally, it of course happens frequently that more than one type of
index appears in an expression, labelling tensors in different
spaces. Consider for instance,
\begin{equation}
C = A^2\quad\text{with}\quad A_{m \mu} = \bar{\psi}\Gamma_{m p} \psi\, B_{p \mu}\,,
\end{equation}
where the roman and Greek indices cannot be interchanged at will,
because they refer to flat and curved spaces respectively. This
example translates to
\begin{screen}
{\mu, \rho, \nu#}::Indices(curved).
{m, n, p, q#}::Indices(flat).
C:= A_{m \nu} A_{m \nu};
@substitute!(
          =  \bar{\psi}\Gamma_{m p} \psi B_{p \mu \rho} C_{\rho});
\end{screen}
with the expected result
\begin{screen}
C:= \bar{\psi} \Gamma_{m p} \psi B_{p \nu \rho} C_{\rho}
    \bar{\psi} \Gamma_{m n} \psi B_{n \nu \mu} C_{\mu};
\end{screen}
All this type of relabelling is done by the internal tree manipulation
algorithms, which ensures that no algorithm can lead to inconsistent
expressions. New dummy indices are taken from the appropriate sets,
using the property information associated to the various indices.

\subsection{Canonicalisation and Young-tableaux methods}

As long as one deals only with symmetric or antisymmetric tensors,
many computer algebra systems are able to write tensor monomials in a
canonical form (although efficient algorithms for very large numbers
of indices or very large numbers of identical tensors have only
surfaced relatively recently,
see~\cite{Portugal:1998qi,port2,e_xact,e_canon}). Generic algorithms
for multi-term Garnir symmetries, such as the Ricci or Bianchi
identity, are much less widespread; see the discussion in
section~\ref{s:symmetries}. Cadabra is the first system to label
tensors by Young tableaux and to use Young projector methods to handle
multi-term symmetries.

A common problem in which multi-term symmetries play an important role
is the construction of a basis of all tensor monomials of a given
length dimension. Determining the number of elements of such a basis
is a relatively straightforward exercise in group
theory~\cite{Fulling:1992vm}. In order to actually construct the
basis, cadabra uses the Young projector method described in the
appendix of~\cite{Green:2005qr}. As an example, let us construct a
basis of monomials cubic in the Riemann tensor,
\begin{screen}
{m,n,p,q,r,s,t,u,v,w,a,b}::Indices(vector).
{m,n,p,q,r,s,t,u,v,w,a,b}::Integer(0..9).
R_{m n p q}::RiemannTensor.

basisR3:= R_{m n p q} R_{r s t u} R_{v w a b};
@all_contractions(

@canonicalise!(
@substitute!(
@substitute!(
\end{screen}
After a declaration of the objects to be used, the program determines
in one step all possible independent contractions of three Riemann
tensors. The last two lines only serve to rewrite the result in terms
of Ricci tensors and scalars, after which the output takes the form
\begin{screen}
basisR3:= \{ R_{m n p q} R_{m p r s} R_{n r q s}, 
             R R_{q r} R_{q r}, 
             R_{n p} R_{n q p r} R_{q r}, 
             R_{n p} R_{n q r s} R_{p r q s}, 
             R R_{p q r s} R_{p q r s}, 
             R_{n p} R_{n r} R_{p r}, 
             R_{m n p q} R_{m r p s} R_{n r q s}, 
             R R R \};
\end{screen}
This result is equivalent to the basis given in the ``${\mathcal R}^0_{6,3}$''
table on page~1184 of~\cite{Fulling:1992vm}.

It is also possible to decompose any given tensor monomial on a
previously constructed basis. Take for example the basis of
Weyl tensor monomials of fourth order. This basis can be read off from
the tables of~\cite{Fulling:1992vm},
\begin{equation}
\begin{aligned}
W_1 &= W_{m n a b} W_{n p b c} W_{p s c d} W_{s m d a}\,,\\[1ex]
W_2 &= W_{m n a b} W_{n p b c} W_{m s c d} W_{s p d a}\,,\\[1ex]
W_3 &= W_{m n a b} W_{p s b a} W_{m n c d} W_{p s d c}\,,\\[1ex]
W_4 &= W_{m n a b} W_{m n b a} W_{p s c d} W_{p s d c}\,,\\[1ex]
W_5 &= W_{m n a b} W_{n p b a} W_{p s c d} W_{s m d c}\,,\\[1ex]
W_6 &= W_{m n a b} W_{p s b a} W_{m p c d} W_{n s d c}\,,\\[1ex]
W_7 &= W_{m n}{}^{[m n} W_{p q}{}^{p q} W_{r s}{}^{r s} W_{t u}{}^{t u]}\,.
\end{aligned}
\end{equation}
If we want to find the decomposition
\begin{equation}
W_{p q r s} W_{p t r u} W_{t v q w} W_{u v s w}- W_{p q r s} W_{p q t u} W_{r v t w} W_{s v u w} 
 = W_2 - \tfrac{1}{4} W_6\,,
\end{equation}
using ``classical'' methods, we would need to figure out the right way
to repeatedly apply the Ricci cyclic identity to the left-hand side of
this expression. The appropriate program to decompose the left-hand
side on the seven-term basis and prove this identity is
\begin{screen}
{m,n,p,q,r,s,t,u,v,w,a,b,c,d,e,f}::Indices(vector).
W_{m n p q}::WeylTensor.

W1:= W_{m n a b} W_{n p b c} W_{p s c d} W_{s m d a};
W2:= W_{m n a b} W_{n p b c} W_{m s c d} W_{s p d a};
W3:= W_{m n a b} W_{p s b a} W_{m n c d} W_{p s d c};
W4:= W_{m n a b} W_{m n b a} W_{p s c d} W_{p s d c};
W5:= W_{m n a b} W_{n p b a} W_{p s c d} W_{s m d c};
W6:= W_{m n a b} W_{p s b a} W_{m p c d} W_{n s d c};
W7:= W_{m n}^{m n} W_{p q}^{p q} W_{r s}^{r s} W_{t u}^{t u};
@asym!(
@substitute!(
@indexsort!(
@collect_terms!(
@canonicalise!(
@collect_terms!(

basisW4:= { @(W1), @(W2), @(W3), @(W4), @(W5), @(W6), @(W7) };

W_{p q r s} W_{p t r u} W_{t v q w} W_{u v s w} 
  - W_{p q r s} W_{p q t u} W_{r v t w} W_{s v u w};
@decompose!(
@list_sum!(
@collect_terms!(
\end{screen}
Most of this code is self-explanatory. The first two lines declare the
  symbols and objects to be used, the next block of lines declares the
  basis and performs the eight-fold anti-symmetrisation for the last
  basis element.\footnote{Commands such as {\tt @collect\_terms} can
  be added to a list of default rules to be applied automatically;
  they have been included here so that all steps are explicit.}
  The decomposition is done with the last three lines. The
  final output of this small program reads
\begin{screen}
{0, 1, 0, 0, 0, -1/4, 0 };
\end{screen}
Internally, this involved a Young-projection of all tensors in the
basis, a projection of the tensors in the expression which we want to
decompose, and a solution of a system of linear
equations~\cite{Green:2005qr}. The internal algorithm is completely
generic and applies to tensor monomials with arbitrary symmetries.

\subsection{Properties and property inheritance}

A typical class of problems in which one handles tensors of both
commuting and anti-commuting type is the construction of
supersymmetric actions. This class of problems also shows the use of
implicit dependence of tensors on coordinates, as well as inheritance
of spinor and anti-commutativity properties. 

Consider as a trivial example -- which is nevertheless not easy to
reproduce with other computer algebra systems -- the invariance of the
super-Maxwell action
\begin{equation}
S = \int\!{\rm d^4}x\, \Big[ -\frac{1}{4} (f_{ab})^2 -
\frac{1}{2}\bar{\lambda}\gamma^a \partial_a \lambda\Big]\,,
\end{equation}
(where~$f_{ab} = \partial_a A_b - \partial_b A_a$) under the
transformations
\begin{equation}
\delta A_a = \bar{\epsilon}\gamma_a \lambda\,,\quad
\delta \lambda = -\frac{1}{2} \gamma^{a b} \epsilon\, f_{a b}\,.
\end{equation}
The object properties for this problem are
\begin{screen}
{ a,b,c,d,e }::Indices(vector).
\bar{#}::DiracBar.
{ \partial{#}, \ppartial{#} }::PartialDerivative.
{ A_{a}, f_{a b} }::Depends(\partial, \ppartial).
{ \epsilon, \gamma_{#} }::Depends(\bar).
\lambda::Depends(\bar, \partial).
{ \lambda, \gamma_{#} }::NonCommuting.
{ \lambda, \epsilon }::Spinor(dimension=4, type=Majorana).
{ \epsilon, \lambda }::SortOrder.
{ \epsilon, \lambda }::AntiCommuting.
\lambda::SelfAntiCommuting.
\gamma_{#}::GammaMatrix.
\delta{#}::Accent.
f_{a b}::AntiSymmetric.
\delta_{a b}::KroneckerDelta.
\end{screen}
Note the use of two types of properties: those which apply to a single
object, like {\tt Depends}, and those which are associated to a list
of objects, like {\tt AntiCommuting}. Clearly $\partial_a \lambda$ and
$\epsilon$ are anti-commuting too, but the program figures this out
automatically from the fact that {\tt $\backslash$partial} has a {\tt
PartialDerivative} property associated to it.

The actual calculation is an almost direct transcription of the
calculation one would do by hand.\footnote{This example makes use of a
set of default rules, to wit ``{\tt ::PostDefaultRules(
@@prodsort!(\%), @@rename\_dummies!(\%), @@canonicalise!(\%),
@@collect\_terms!(\%) )}'', which mimick the automatic rewriting
behaviour of many other computer algebra systems and get invoked
automatically at each step. See~\cite{kas_cdb} for more details.}
First we define the supersymmetry transformation rules and the action,
which can be entered as in~\TeX{},
\begin{screen}
susy:= { \delta{A_{a}}   = \bar{\epsilon} \gamma_{a} \lambda, 
         \delta{\lambda} = -(1/2) \gamma_{a b} \epsilon f_{a b} };

S:= -(1/4) f_{a b} f_{a b} 
              - (1/2) \bar{\lambda} \gamma_{a} \partial_{a}{\lambda};
\end{screen}
Showing invariance starts by applying a variational derivative,
\begin{screen}
@vary!(
           \lambda -> \delta{\lambda} );

@distribute!(
@substitute!(
\end{screen}
After these steps, the result is (shown exactly as it appears in the
graphical and the \TeX{}macs~\cite{vdH:Gut} front-ends)%
\begin{equation}
S = \bar{\epsilon} \gamma_{a} \partial_{b} \lambda\, f_{ab}
  + \frac{1}{4} \overline{\gamma_{ab} \epsilon} \gamma_{c}
  \partial_c\lambda\, f_{ab}
  + \frac{1}{4} \bar{\lambda}\gamma_a \gamma_{bc} \epsilon \partial_a f_{bc}\,.
\end{equation}
Since the program knows about the properties of gamma matrices it can
rewrite the Dirac bar, and then we do one further partial integration,
\begin{screen}
@rewrite_diracbar!(
@substitute!(
@pintegrate!(
@rename!(
@prodrule!(
\end{screen}
What remains is the gamma matrix algebra, a rewriting of the
derivative of the Dirac bar as the Dirac bar of a derivative, and
sorting of spinors (which employs inheritance of the {\tt Spinor} and
{\tt AntiCommuting} properties as already alluded to earlier),
\begin{screen}
@join!(
@substitute!(
@spinorsort!(
\end{screen}
The result is (after partial integration) a Bianchi identity on the
field strength, and thus invariance of the action.

While this example is rather simple, and does not require a computer
algebra system for its solution, it illustrates that the extended tree
structure together with the property system make it possible to
manipulate expressions in a way which closely resembles what one would
do when solving the problem with pencil and paper. Several more
complicated examples will be discussed in the
upcoming~\cite{kas_cdb_hep}.

\section{Summary}

I have presented a new prototype computer algebra system which is
designed to be an easy-to-use scratch pad for problems encountered in
field theory. The current library of algorithms include functionality
to deal with bosonic and fermionic tensors, spinors, gamma matrices,
differential operators and so on, all through the use of a
multiple-inheritance property mechanism. Cadabra is the first system
which handles generic multi-term tensor symmetries using a
Young-projector based algorithm. It is also the first system which
accepts input in~\TeX{} form, eliminating tedious translation steps
and making programs much easier to read for new users. Finally, the
source code of the system is freely available and the reference guide
contains extensive documentation explaining how to add new algorithm
modules to the program.

\section*{Acknowledgements}

I am grateful to Jos\'e Martin-Garcia for inspiring discussions and
for help with the use of his {\tt xPerm} code~\cite{e_xact} for
mono-term canonicalisation. I thank the anonymous referee for
extensive comments which have substantially improved this paper.

\setlength{\bibsep}{4pt}
\begingroup\raggedright\endgroup

\end{document}